\documentstyle[12pt]{article}

\begin{document}

\begin{titlepage}{\LARGE
\begin{center} Elliptic Ruijsenaars--Schneider\\
and Calogero--Moser hierarchies\\
are governed by the same $r$--matrix \end{center}}

\vspace{1.5cm}

\begin{flushleft}{\large Yuri B. SURIS}\end{flushleft} \vspace{1.0cm}
Centre for Complex Systems and Visualization, University of Bremen,\\
Postfach 330 440, 28334 Bremen, Germany\\
e-mail: suris @ mathematik.uni-bremen.de 

\vspace{2.0cm}

{\small {\bf Abstract.} We demonstrate that in a certain gauge the elliptic
Ruijsenaars--Schneider models admit Lax representation governed
by the same dynamical $r$--matrix as their non--relativistic counterparts 
(Calogero--Moser models). This phenomenon was previously observed for the 
rational and hyperbolic models.}
\end{titlepage}

\setcounter{equation}{0}
\section{Introduction}

In the recent years the interest in the Calogero--Moser type of models
\cite{OP}--\cite{R} is considerably revitalized. One of the directions of 
this recent development was connected with the notion of the dynamical 
$r$--matrices and their interpretation in terms of Hamiltonian reduction 
\cite{AT}--\cite{N}. Very recently \cite{BB},\cite{AR}, \cite{S}, \cite{N}, 
this line of research included also the so--called Ruijsenaars--Schneider 
models \cite{RS},\cite{R} which may be seen as relativistic generalizations 
of the Calogero--Moser ones \cite{OP}, \cite{KKS}. 

In the paper \cite{BB} a {\it quadratic} (dynamical) $r$--matrix Poisson
structure was found for the dynamical system describing the motion of the 
solitons of the sine--Gordon model. This system turns out to be a particular 
case of the hyperbolic Ruijsenaars--Schneider model corresponding to a 
particular value of the parameter $\gamma$ of the model (cf. (\ref{L rel}) 
below), namely $\gamma=i\pi/2$, when the Lax matrix becomes symmetric in some
gauge, 

The case of general hyperbolic Ruijsenaars--Schneider model, 
admitting also the rational model as a limiting
case, was considered in \cite{AR}. There was found a {\it linear} $r$--matrix
structure for this model, with the linear dependence of the $r$--matrix on the
elements of the Lax matrix. However, as it stands, the structure found in 
\cite{AR} cannot be cast into a quadratic form. 

This drawback was overcome in \cite{S}, where the quadratic $r$--matrix Poisson 
bracket was found for the general rational and hyperbolic models. Moreover, this 
bracket turned out to posess several remarkable properties. 
\begin{itemize}
\item
First, the $r$--matrix objects turned out to be independent on the 
relativistic parameter $\gamma$ of the model. 
\item
Second, and more important, the $r$--matrix 
object governing the whole hierarchy of the Lax equations attached to
the Ruijsenaars--Schneider model turned out to be
{\it identical} with the corresponding object governing the 
non--relativistic Calogero--Moser hierarchy. 
\end{itemize}
A geometric interpretation of this
intriguing property was also provided in \cite{S}. Several open problems were 
formulated in \cite{S}, the first of them being the generalization of these 
findings to the case of elliptic Ruijsenaars--Schneider model.

Soon after \cite{S} there appeared the paper \cite{N} where an $r$--matrix 
quadratic Poisson bracket for the elliptic Ruijsenaars--Schneider model was 
found, thus partly solving the mentioned problem. However, despite
the fact that this bracket has the same general structure as the one found in 
\cite{S}, it turns out not to generalize  the latter. It fails to 
have the two remarkable properties pointed out above, and moreover it does not 
reduce to the bracket found in \cite{S} in the corresponding (rational or
hyperbolic) limit.

In the present paper we give a proper generalization of the results in 
\cite{S} for the elliptic case. Namely, we present a quadratic $r$--matrix 
structure for this model enjoying the two properties listed above. An existence 
of two different $r$--matrix Poisson brackets for one and the same model is
not contradictory, because of the well--known non--uniqueness of an $r$--matrix.
So in principle both can coexist on their own rights. We hope, however,
that the two remarkable properties pointed out above indicate on some deeper
geometric meaning to be clarified in the future, so
that the result reported here will be accepted as {\it the} $r$--matrix for the
elliptic Ruijsenaars--Schneider model.

\setcounter{equation}{0}
\section{Elliptic models of the Calogero--Moser type.}

The elliptic non--relativistic Calogero--Moser hierarchy is described in
terms of the {\it Lax matrix}
\begin{equation}\label{L nr}
L=L(x,p,\lambda)=\sum_{k=1}^Np_kE_{kk}+\gamma\sum_{k\neq j}\Phi(x_k-x_j,\lambda)E_{kj}.
\end{equation}
Here the function $\Phi(x,\lambda)$ is defined as 
\begin{equation}\label{Phi}
\Phi(x,\lambda)=\frac{\sigma(x+\lambda)}{\sigma(x)\sigma(\lambda)},
\end{equation}
where $\sigma(x)$ is the Weierstrass $\sigma$--function. Further,
$\lambda$ is an auxiliary (so called spectral) parameter which
does not enter the equations of motion of the model, but rather serves as a
useful tool for its solution. On the contrary, $\gamma$ is an internal  
parameter of the model, usually supposed to be pure imaginary. 
The dynamical variables $x=(x_1,\ldots,x_N)^T$ and $p=(p_1,\ldots,p_N)^T$ 
are supposed to be canonically conjugated, i.e. to have canonical Poisson 
brackets:
\begin{equation}\label{can PB}
\{x_k,x_j\}=\{p_k,p_j\}=0,\quad \{x_k,p_j\}=\delta_{kj}.
\end{equation}
The Hamiltonian function of the Calogero--Moser model proper (i.e. of the 
simplest representative of the Calogero--Moser hierarchy) is given by
\[
H(x,p)=\frac{1}{2}\sum_{k=1}^N p_k^2-
\frac{1}{2}\gamma^2\sum_{k\neq j}\wp(x_k-x_j)=\frac{1}{2}{\rm tr}L^2(x,p,\lambda)
+{\rm const},
\]
where ${\rm const}=-N(N-1)\gamma^2\wp(\lambda)/2$, and $\wp(x)$ is the
Weierstrass elliptic function.

The elliptic relativistic Ruijsenaars--Schneider hierarchy is also described 
in terms of the {\it Lax matrix}
\begin{equation}\label{L rel}
L(x,p,\lambda)=\sum_{k,j=1}^N \frac{\Phi(x_k-x_j+\gamma,\lambda)}
{\Phi(\gamma,\lambda)}b_jE_{kj}.
\end{equation}
The notations are the same as above, and we use an additional abbreviation:
\begin{equation}\label{def b}
b_k=\exp(p_k)\prod_{j\neq k}\left(
\frac{\sigma(x_k-x_j+\gamma)\sigma(x_k-x_j-\gamma)}{\sigma^2(x_k-x_j)}
\right)^{1/2},
\end{equation}
so that in the variables $(x,b)$ the canonical Poisson brackets (\ref{can PB}) 
take the form
\[
\{x_k,x_j\}=0,\quad \{x_k,b_j\}=b_k\delta_{kj},
\]
\begin{equation}\label{rel PB}
\{b_k,b_j\}=b_kb_j\Big(\zeta(x_j-x_k+\gamma)-
\zeta(x_k-x_j+\gamma)+2(1-\delta_{kj})\zeta(x_k-x_j)\Big).
\end{equation}
Here $\zeta(x)$ is, of course, the Weierstrass $\zeta$--function, i.e.
\[
\zeta(x)=\frac{\sigma'(x)}{\sigma(x)}.
\]
The Hamiltonian function of the Ruijsenaars--Schneider model proper (i.e. of
the simplest member of this hierarchy) is simply
\[
H(x,p)=\sum_{k=1}^N b_k={\rm tr}L(x,p,\lambda).
\]

Let us note that the non--relativistic limit, leading from the 
Ruijsenaars--Schneider model to the Calogero--Moser one, is achieved
by rescaling $p\mapsto\beta p$, $\gamma\mapsto\beta\gamma$ and 
subsequent sending $\beta\to 0$ (in this limit $L_{{\rm rel}}=
I+\beta L_{{\rm nonrel}}+O(\beta^2)$).

Let us also note that the evolution of either of the non--relativistic or the
relativistic model is governed by the {\it Lax equation} of the form
\begin{equation}\label{Lax eq}
\dot{L}=[M,L],
\end{equation}
where, for example, for the Ruijsenaars--Schneider model one has:
\begin{equation}\label{M kj}
M_{kj}=\Phi(x_k-x_j,\lambda)b_j,\;\;k\neq j,
\end{equation}
\begin{equation}\label{M kk}
M_{kk}=\Big(\zeta(\lambda)+\zeta(\gamma)\Big)b_k+
\sum_{j\neq k}\Big(\zeta(x_k-x_j+\gamma)-\zeta(x_k-x_j)\Big)b_j.
\end{equation}
An $r$--matrix found below enables one to give a general formula for the matrix
$M$ for an arbitrary flow of the corresponding hierarchy (cf. \cite{S} for such
formulas in the rational and hyperbolic cases).

\setcounter{equation}{0}
\section{Dynamical $r$-matrix formulation}

An $r$--matrix formulation of the elliptic Calogero--Moser  model was given in
\cite{Skl}, \cite{BS} as a generalization of the previous result obtained in
\cite{AT} for the rational and hyperbolic cases. The result of \cite{Skl} may
be presented in the following form: for the non--relativistic case the 
corresponding Lax matrices satisfy a  linear $r$--matrix ansatz
\begin{equation}\label{r Anz}
\{L(\lambda)\stackrel{\otimes}{,}L(\mu)\}=
\left[I\otimes L(\mu),r(\lambda,\mu)\right]-\left[L(\lambda)\otimes I,
r^*(\lambda,\mu)\right],
\end{equation}
where the $N^2\times N^2$ matrix $r(\lambda,\mu)$ may be decomposed into the sum 
\begin{equation}\label{ras}
r(\lambda,\mu)=a(\lambda,\mu)+s(\lambda).
\end{equation}
Here $a$ is a skew--symmetric matrix 
\begin{equation}\label{a}
a(\lambda,\mu)=-\zeta(\lambda-\mu)\sum_{k=1}^N E_{kk}\otimes E_{kk}-
\sum_{k\neq j}\Phi(x_j-x_k,\lambda-\mu)E_{jk}\otimes E_{kj},
\end{equation}
and $s$ is a non--skew--symmetric one:
\begin{equation}\label{s}
s(\lambda)=\zeta(\lambda)\sum_{k=1}^N E_{kk}\otimes E_{kk}+
\sum_{k\neq j}\Phi(x_j-x_k,\lambda)E_{jk}\otimes E_{kk}.
\end{equation}
Here the ''skew--symmetry'' is understood with respect to the operation
\[
r^*(\lambda,\mu)=\Pi r(\mu,\lambda)\Pi\;\;{\rm with}\;\;
\Pi=\sum_{k,j=1}^NE_{kj}\otimes E_{jk}.
\]
So we have
\[
a^*(\lambda,\mu)=-a(\lambda,\mu),
\]
and
\begin{equation}\label{s*}
s^*(\mu)=\zeta(\mu)\sum_{k=1}^N E_{kk}\otimes E_{kk}+
\sum_{k\neq j}\Phi(x_j-x_k,\mu)E_{kk}\otimes E_{jk}.
\end{equation}

(Note that our $r$ is related to the objects $r_{12}$, $r_{21}$ in \cite{Skl}
by means of $r=-r_{21}$ and $r^*=-r_{12}$).

We shall prove that in the relativistic case the corresponding
Lax matrices satisfy the quadratic $r$--matrix ansatz:
\begin{eqnarray}
\{L(\lambda)\stackrel{\otimes}{,}L(\mu)\} & = &
(L(\lambda)\otimes L(\mu))a_1(\lambda,\mu)-
a_2(\lambda,\mu)(L(\lambda)\otimes L(\mu))\nonumber\\
 & + & (I\otimes L(\mu))s_1(\lambda,\mu)(L(\mu)\otimes I)-
(L(\lambda)\otimes I)s_2(\lambda,\mu)(I\otimes L(\mu))\nonumber\\
 &   &\label{as Anz}
\end{eqnarray}
where the matrices $a_1,a_2,s_1,s_2$ satisfy the conditions
\begin{equation}\label{sym}
a_1^*(\lambda,\mu)=-a_1(\lambda,\mu),\quad a_2^*(\lambda,\mu)=-a_2(\lambda,\mu),
\quad s_2^*(\lambda,\mu)=s_1(\lambda,\mu),
\end{equation}
and
\begin{equation}\label{sum}
a_1(\lambda,\mu)+s_1(\lambda,\mu)=a_2(\lambda,\mu)+s_2(\lambda,\mu)
=r(\lambda,\mu).
\end{equation}
The first of these conditions assures the skew--symmetry of the Poisson bracket
(\ref{as Anz}), and the second one garantees that the Hamiltonian flows with
invariant Hamiltonian functions $\varphi(L)$ have the Lax form (\ref{Lax eq})
with the form of the $M$--matrix being governed by the same $r$--matrix as in
the non--relativistic case.

Such general quadratic $r$-matrix structures were discovered several times 
independently \cite{FM}, \cite{P}, \cite{S2}. See \cite{S2} for an application
to closely related, but much more simple systems of the Toda lattice type.

{\bf Theorem.} {\it For the Lax matrices of the relativistic model 
{\rm (\ref{L rel})} there holds a quadratic $r$--matrix ansatz 
{\rm (\ref{as Anz})} with the matrices
\[
a_1=a+w,\quad s_1=s-w,
\]
\[
a_2=a+s-s^*-w,\quad s_2=s^*+w,
\]
and $w$ is an auxiliary matrix}
\begin{equation}\label{w}
w=\sum_{k\neq j}\zeta(x_k-x_j)E_{kk}\otimes E_{jj}.
\end{equation}

Note that all the objects $a$, $s$, $w$ entering these formula do not depend on
$\gamma$, and that (\ref{sum}) is fulfilled, which justifies the title of the
present paper.

The {\bf proof} of this Theorem is based on direct computations, presented in 
the Appendix.

\setcounter{equation}{0}
\section{Conclusions}

Now that the formal part of the results in \cite{S} is generalized, it is
tempting to find a geometrical explanation of the phenomena behind it. To
this end one should develop further the theory of Calogero--Moser type models as
Hamiltonian reduced systems.

Certainly, this problem should be supplied with the whole list of open problems
formulated in \cite{S}, \cite{N}.

\setcounter{equation}{0}
\section{Acknowledgements}

The research of the author is financially supported by the DFG (Deutsche
Forschungsgemeinschaft). My pleasant duty is to thank warmly Professor
Orlando Ragnisco (University of Rome) for organizing my visit to Rome, where
this work was done, for useful discussions and collaboration, and his
institution -- for financial support during this visit.

\setcounter{equation}{0}
\section{Appendix: proof of the Theorem}

Let us denote
\[
\{L_{ij}(\lambda),L_{km}(\mu)\}=\pi_{ijkm}L_{ij}(\lambda)L_{km}(\mu),
\]
\[
[L(\lambda)\otimes L(\mu),\,a(\lambda,\mu)]=
\sum_{i,j,k,m=1}^N \alpha_{ijkm}L_{ij}(\lambda)L_{km}(\mu)E_{ij}\otimes E_{km},
\]
and analogously
\[
s(\lambda)(L(\lambda)\otimes L(\mu))=
\sum_{i,j,k,m=1}^N \sigma^{(1)}_{ijkm}L_{ij}(\lambda)L_{km}(\mu)E_{ij}\otimes 
E_{km},
\]
\[
s^*(\mu)(L(\lambda)\otimes L(\mu))=
\sum_{i,j,k,m=1}^N \sigma^{(2)}_{ijkm}L_{ij}(\lambda)L_{km}(\mu)E_{ij}\otimes 
E_{km},
\]
\[
(I\otimes L(\mu))s(\lambda)(L(\lambda)\otimes I)=
\sum_{i,j,k,m=1}^N \sigma^{(3)}_{ijkm}L_{ij}(\lambda)L_{km}(\mu)E_{ij}\otimes 
E_{km},
\]
\[
(L(\lambda)\otimes I)s^*(\mu)(I\otimes L(\mu))=
\sum_{i,j,k,m=1}^N \sigma^{(4)}_{ijkm}L_{ij}(\lambda)L_{km}(\mu)E_{ij}\otimes 
E_{km}.
\]
The statement of the Theorem is equivalent to
\begin{equation}\label{form}
\pi_{ijkm}=\alpha_{ijkm}-\sigma^{(1)}_{ijkm}
+\sigma^{(2)}_{ijkm}+\sigma^{(3)}_{ijkm}-\sigma^{(4)}_{ijkm}+w_{ik}
+w_{jm}-w_{im}-w_{jk},
\end{equation}
where $w_{jk}=(1-\delta_{jk})\zeta(x_j-x_k)$ are the coefficients of the auxiliary
matrix $w=\sum_{j\neq k}w_{jk}E_{jj}\otimes E_{kk}$.

According to the Poisson brackets (\ref{rel PB}) we have:
\[
\pi_{ijkm}=\zeta(x_m-x_j+\gamma)-
\zeta(x_j-x_m+\gamma)+2(1-\delta_{jm})\zeta(x_j-x_m)
\]
\[
+(\delta_{jk}-\delta_{jm})
\Big(\zeta(x_k-x_m+\gamma)-\zeta(x_k-x_m+\gamma+\mu)\Big)
\]
\begin{equation}\label{pi}
-(\delta_{im}-\delta_{jm})
\Big(\zeta(x_i-x_j+\gamma)-\zeta(x_i-x_j+\gamma+\lambda)\Big).
\end{equation}

From the definitions of the matrices $a$ and $s$ we have:
\begin{eqnarray*}
\alpha_{ijkm}
  & = & (\delta_{ik}-\delta_{jm})\zeta(\lambda-\mu)\\
  &   & + (1-\delta_{ik})\frac{L_{kj}(\lambda)L_{im}(\mu)}
        {L_{ij}(\lambda)L_{km}(\mu)}\Phi(x_i-x_k,\lambda-\mu)\\
  &   & - (1-\delta_{jm})\frac{L_{im}(\lambda)L_{kj}(\mu)}
        {L_{ij}(\lambda)L_{km}(\mu)}\Phi(x_m-x_j,\lambda-\mu);\\ \\
\sigma^{(1)}_{ijkm}
  & = & \delta_{ik}\zeta(\lambda)+
        (1-\delta_{ik})\frac{L_{kj}(\lambda)}{L_{ij}(\lambda)}
        \Phi(x_i-x_k,\lambda);\\ \\
\sigma^{(2)}_{ijkm}
  & = & \delta_{ik}\zeta(\mu)+
        (1-\delta_{ik})\frac{L_{im}(\mu)}{L_{km}(\mu)}
        \Phi(x_k-x_i,\mu);\\ \\
\sigma^{(3)}_{ijkm}
  & = & \delta_{im}\zeta(\lambda)+
        (1-\delta_{im})\frac{L_{mj}(\lambda)}{L_{ij}(\lambda)}
        \Phi(x_i-x_m,\lambda);\\ \\
\sigma^{(4)}_{ijkm}
  & = & \delta_{jk}\zeta(\mu)+
       (1-\delta_{jk})\frac{L_{jm}(\mu)}{L_{km}(\mu)}
        \Phi(x_k-x_j,\mu).\\ \\
\end{eqnarray*}

Using the expressions for the elements of the matrix $L$, we get:

\begin{eqnarray*}
\alpha_{ijkm}
  & = & (\delta_{ik}-\delta_{jm})\zeta(\lambda-\mu)\\
  &   & + (1-\delta_{ik})\frac{\Phi(x_k-x_j+\gamma,\lambda)
        \Phi(x_i-x_m+\gamma,\mu)\Phi(x_i-x_k,\lambda-\mu)}
       {\Phi(x_i-x_j+\gamma,\lambda)\Phi(x_k-x_m+\gamma,\mu)};\\ 
  &   & -(1-\delta_{jm})\frac{\Phi(x_i-x_m+\gamma,\lambda)
         \Phi(x_k-x_j+\gamma,\mu)\Phi(x_m-x_j,\lambda-\mu)}
        {\Phi(x_i-x_j+\gamma,\lambda)\Phi(x_k-x_m+\gamma,\mu)};\\ \\
\sigma^{(1)}_{ijkm}
  & = & \delta_{ik}\zeta(\lambda)+
        (1-\delta_{ik})\frac{\Phi(x_k-x_j+\gamma,\lambda)\Phi(x_i-x_k,\lambda)}
        {\Phi(x_i-x_j+\gamma,\lambda)};\\ \\
\sigma^{(2)}_{ijkm}
  & = & \delta_{ik}\zeta(\mu)+
        (1-\delta_{ik})\frac{\Phi(x_i-x_m+\gamma,\mu)\Phi(x_k-x_i,\mu)}
        {\Phi(x_k-x_m+\gamma,\mu)};\\ \\
\sigma^{(3)}_{ijkm}
  & = & \delta_{im}\zeta(\lambda)+
        (1-\delta_{im})\frac{\Phi(x_m-x_j+\gamma,\lambda)\Phi(x_i-x_m,\lambda)}
        {\Phi(x_i-x_j+\gamma,\lambda)};\\ \\
\sigma^{(4)}_{ijkm}
  & = & \delta_{jk}\zeta(\mu)+
        (1-\delta_{jk})\frac{\Phi(x_j-x_m+\gamma,\mu)\Phi(x_k-x_j,\mu)}
        {\Phi(x_k-x_m+\gamma,\mu)}.
\end{eqnarray*}

The most laburous part of the further manipulations is the simplification of
the expression for $\alpha_{ijkm}$. This was performed already in \cite{N}, we
give here slightly more details. Following two elliptic identities were used
in \cite{N} to this aim:
\[
\frac{\Phi(X-A,\lambda)\Phi(Y+A,\mu)\Phi(A,\lambda-\mu)-
\Phi(Y+A,\lambda)\Phi(X-A,\mu)\Phi(X-Y+A,\lambda-\mu)}
{\Phi(X,\lambda)\Phi(Y,\mu)}=
\]
\[
\zeta(A)-\zeta(X-Y+A)+\zeta(X-A)-\zeta(Y+A),
\]
and
\[
\frac{\Phi(Y,\lambda)\Phi(X,\mu)\Phi(X-Y,\lambda-\mu)}
{\Phi(X,\lambda)\Phi(Y,\mu)}=
\zeta(\lambda-\mu)+\zeta(X-Y)-\zeta(X+\lambda)+\zeta(Y+\mu).
\]
One gets:
\[
\alpha_{ijkm}=(\delta_{ik}-\delta_{jm})\zeta(\lambda-\mu)
\]
\[
+(1-\delta_{ik})(1-\delta_{jm})\Big(\zeta(x_i-x_k)-\zeta(x_m-x_j)
+\zeta(x_k-x_j+\gamma)-\zeta(x_i-x_m+\gamma)\Big)
\]
\[
+(1-\delta_{ik})\delta_{jm}\Big(\zeta(\lambda-\mu)+\zeta(x_i-x_k)
-\zeta(x_i-x_j+\gamma+\lambda)+\zeta(x_k-x_m+\gamma+\mu)\Big)
\]
\[
-(1-\delta_{jm})\delta_{ik}\Big(\zeta(\lambda-\mu)+\zeta(x_m-x_j)
-\zeta(x_i-x_j+\gamma+\lambda)+\zeta(x_k-x_m+\gamma+\mu)\Big).
\]
Further straightforward manipulations give:
\[
\alpha_{ijkm}=(1-\delta_{ik})\zeta(x_i-x_k)-(1-\delta_{jm})\zeta(x_m-x_j)
\]
\[
+\zeta(x_k-x_j+\gamma)-\zeta(x_i-x_m+\gamma)
\]
\[
+(\delta_{ik}-\delta_{jm})\Big(\zeta(x_k-x_m+\gamma)-
\zeta(x_k-x_m+\gamma+\mu)\Big)
\]
\begin{equation}\label{alfa}
-(\delta_{ik}-\delta_{jm})\Big(\zeta(x_i-x_j+\gamma)-
\zeta(x_i-x_j+\gamma+\lambda)\Big).
\end{equation}

By simplifying the expressions for $\sigma_{ijkm}^{(1-4)}$ one uses 
systematically the identity
\[
\frac{\Phi(X,\lambda)\Phi(Y,\lambda)}{\Phi(X+Y,\lambda)}=
\zeta(\lambda)+\zeta(X)+\zeta(Y)-\zeta(X+Y+\lambda).
\]

One gets following expressions:
\begin{eqnarray*}
\sigma^{(1)}_{ijkm}
  & = & \zeta(\lambda)+(1-\delta_{ik})\Big(\zeta(x_i-x_k)+\zeta(x_k-x_j+\gamma)-
         \zeta(x_i-x_j+\gamma+\lambda)\Big);\\ \\
\sigma^{(2)}_{ijkm}
  & = & \zeta(\mu)+(1-\delta_{ik})\Big(\zeta(x_k-x_i)+\zeta(x_i-x_m+\gamma)-
         \zeta(x_k-x_m+\gamma+\mu)\Big);\\ \\
\sigma^{(3)}_{ijkm}
  & = & \zeta(\lambda)+(1-\delta_{im})\Big(\zeta(x_i-x_m)+\zeta(x_m-x_j+\gamma)-
         \zeta(x_i-x_j+\gamma+\lambda)\Big);\\ \\
\sigma^{(4)}_{ijkm}
  & = & \zeta(\mu)+(1-\delta_{jk})\Big(\zeta(x_k-x_j)+\zeta(x_j-x_m+\gamma)-
         \zeta(x_k-x_m+\gamma+\mu)\Big).
\end{eqnarray*}

It follows after straightforward manipulations:
\[
-\sigma_{ijkm}^{(1)}+\sigma_{ijkm}^{(2)}+\sigma_{ijkm}^{(3)}
-\sigma_{ijkm}^{(4)}=
\]
\[
2(1-\delta_{ik})\zeta(x_k-x_i)+(1-\delta_{im})\zeta(x_i-x_m)-
(1-\delta_{jk})\zeta(x_k-x_j)
\]
\[
+\zeta(x_i-x_m+\gamma)-\zeta(x_k-x_j+\gamma)+\zeta(x_m-x_j+\gamma)
-\zeta(x_j-x_m=\gamma)
\]
\[
+(\delta_{ik}-\delta_{im})\Big(\zeta(x_i-x_j+\gamma)-\zeta(x_i-x_j+\gamma
+\lambda)\Big)
\]
\begin{equation}\label{sigmas}
-(\delta_{ik}-\delta_{jk})\Big(\zeta(x_k-x_m+\gamma)-\zeta(x_k-x_m+\gamma
+\mu)\Big).
\end{equation}

Now it is easy to see that combining (\ref{alfa}), (\ref{sigmas}), and
(\ref{pi}), one gets (\ref{form}), which proves the Theorem.

\end{document}